\documentclass{article}

     \usepackage[final]{neurips_2024}

\usepackage[utf8]{inputenc} 
\usepackage[T1]{fontenc}    
\usepackage{hyperref}       
\usepackage{url}            
\usepackage{booktabs}       

\usepackage{amsfonts}       
\usepackage{nicefrac}       
\usepackage{microtype}      
\usepackage{xcolor}         

\usepackage{graphicx}

\setcitestyle{numbers,square}

\title{Creative Loss: Ambiguity, Uncertainty and Indeterminacy}

\author{%
  Tom Holberton\thanks{Associate Professor Unit 21 BSc MArch Architecture - https://unit-21.com} \\
  Bartlett School of Architecture\\
  University College London\\
  London WC1H 0QB\\
  \texttt{t.holberton@ucl.ac.uk} \\
}

\begin{document}

\maketitle

\begin{abstract}
This article evaluates how creative uses of machine learning can address three adjacent terms: \textit{ambiguity}, \textit{uncertainty} and \textit{indeterminacy}.  Through the progression of these concepts it reflects on increasing ambitions for machine learning as a creative partner, illustrated with research from Unit 21 at the Bartlett School of Architecture, UCL.  Through \textit{indeterminacy} are potential future approaches to machine learning and design. 
\end{abstract}

\section{Introduction}

\textit{Ambiguity}, \textit{uncertainty} and \textit{indeterminacy} can strengthen creative work, feeding a sense of the unfamiliar and original. Yet they operate in different realms: from phenomenology to epistemology to ontology —  comparable to the mapped, the navigable and the uncharted \cite{ciprut_indeterminacy_2009}. These differences have significant implications for the role machine learning might play in relation to design, data and agency.  

This article presents works from the research of Unit 21 at the Bartlett School of Architecture \cite{holberton_unit_2024}. It evaluates a number of original design processes developed for specific projects and contexts. Rather than creating generalised models of architectural intelligence, we interrogate the latent space as a new space that can be trained and used differently for each project.

\section{Ambiguity}

To create with \textit{ambiguity} is to tease. It is to anticipate and defy the expectations of the audience through an obfuscation of intention, through the  multiplication of meaning.

Out of a history of classifying objects came machine vision, dreaming \cite{mordvintsev_inceptionism_2015} and hallucinations. As with associated Surrealism this comes with a kind of \textit{over-determination} of meaning, from an abundance of links and ties \cite{conde_architecture_2000}. This morphing and hybrid space is an ambiguous in-between, where \textit{a~priori} category boundaries become overlapping. Ambiguity implicitly accepts the blackboxed model, where the blurred outputs are always the products of an input data with predetermined structure and cultural topologies \cite{lury_introduction_2012}.   Through a spatialisation of this pre-determined order, we can then map a new polyvalent territory. Typically this serves to reinforce the corpus, described by Carpo~\cite{carpo_imitation_2023} as imitation, producing iterations and transfers of style and history \cite{del_campo_towards_2021} \cite{klingemann_memories_2018}.

\newpage
In \textit{\textbf{AI Wabi-sabi}} \cite{holberton_last_2021}\cite{holberton_sea_2021}\cite{holberton_stud_2021}\cite{holberton_twitching_2021}\cite{holberton_glory_2022}\cite{holberton_loyal_2022}, a customised conditional GAN \cite{isola_image--image_2018} is trained on the correspondence between form and texture of traditional Japanese tea bowls. The latent space is treated as ambiguous and amorphous, shaped by the cumulative sense of imperfection in thousands of examples of this craft, which defies overt classification. New bowls are then created from a satellite image (Figure 1 and 2) - an Anthropocene sampling of the earth at one precise location and time which is transitory and impermanent. The GAN and code converts squares of ‘digital clay’ into new 3D forms and sculptural physical bowls tied to that moment.

  \begin{figure}[h!]
 \centering
  \includegraphics[width=1\textwidth]{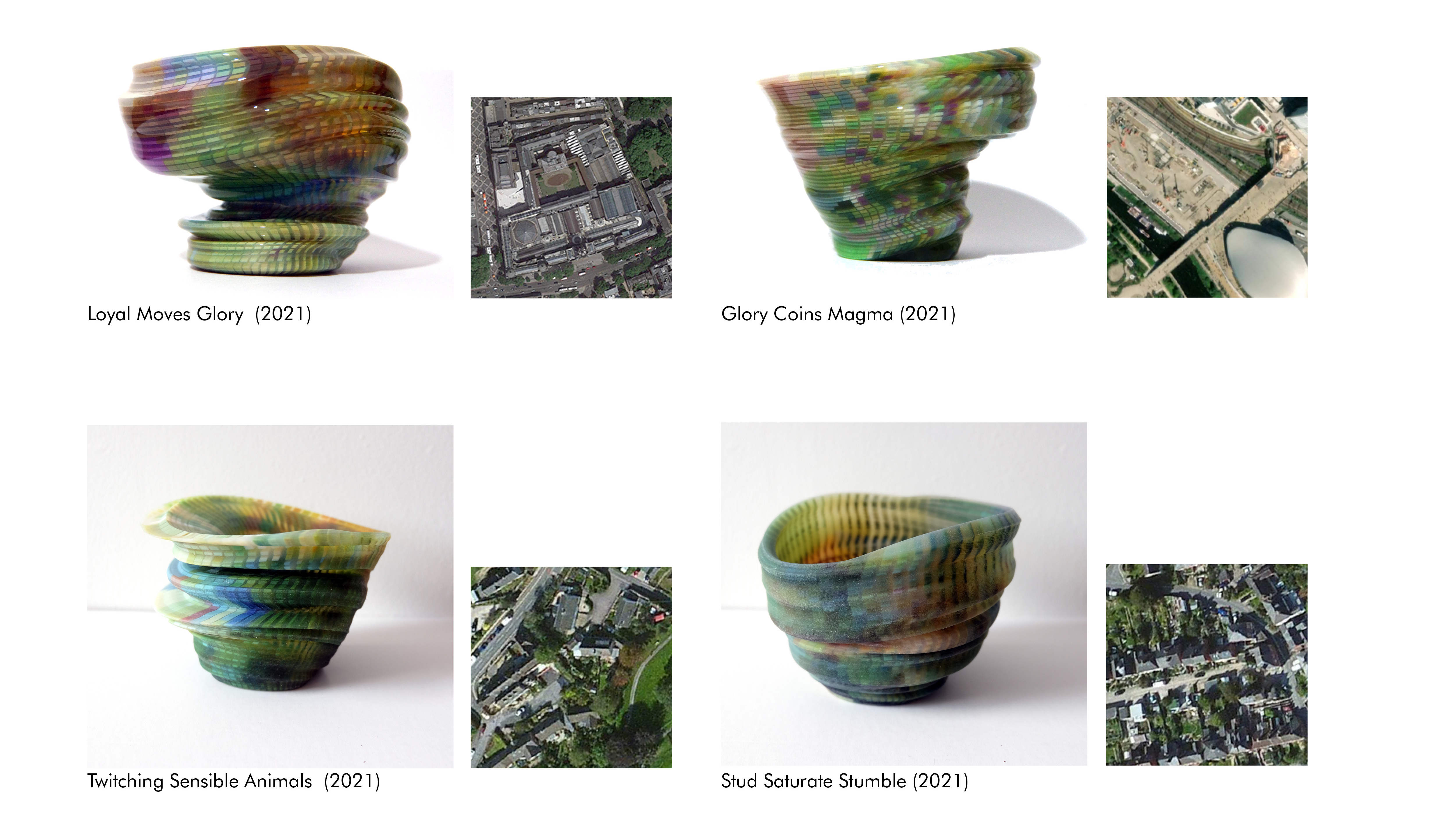}
 \caption{\textit{AI wabi-sabi} : Four physical bowls generated from the 'digital clay' of satellite images and trained from a craft of imperfection.}
 \end{figure}
 
   \begin{figure}[h!]
 \centering
  \includegraphics[width=1\textwidth]{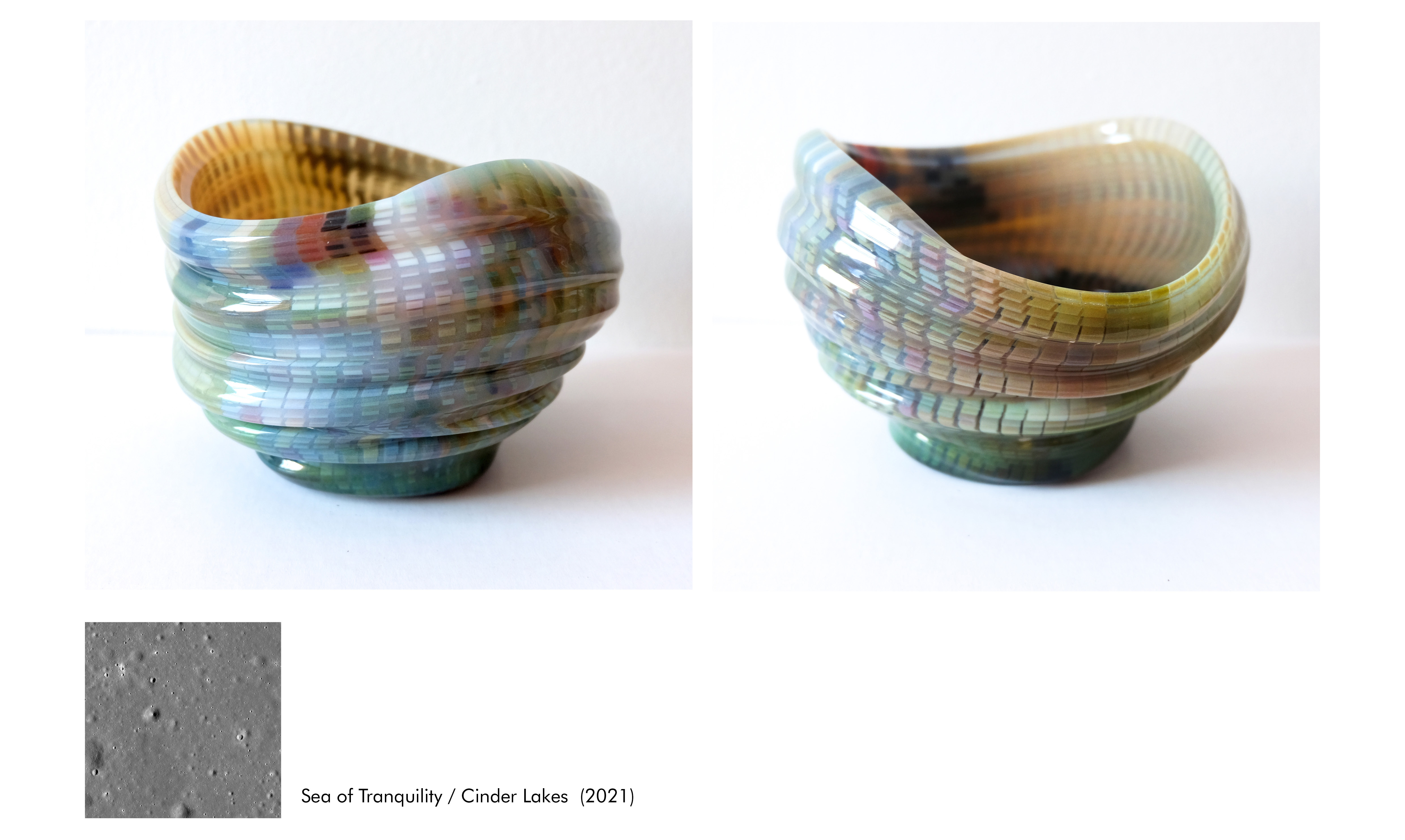}
 \caption{\textit{AI wabi-sabi} : Acrylic and hand lacquered bowl generated from lunar site.}
 \end{figure}

\newpage
\section{Uncertainty}

To create with \textit{uncertainty} is to operate probabilistically.   

In transformers \cite{openai_gpt-4_2023} the next probable word, at the ideal ‘temperature’, offers a thermodynamic notion of creativity. Discrete atomic fragments of words or images ‘bubble away’ with differing probabilities, the sum of the vectors elevating a few favourites to escape this Brownian cloud to be the next item in the sequence. Outcomes are not ambiguous, instead each answer is less or more probable within an overall system of certainties.

Creativity emerges from these \textit{open works} through a sequential expectation.  Umberto Eco stated contemporary art constantly challenges the initial order by means of an extremely “improbable” form of organization \cite{eco_open_1989}. Uncertainty negotiates between the degree to which a creative work maintains openness and dynamism, versus the “structural vitality” that comes from more probable forms rather than a purely random sequence \cite{eco_open_1989}. In architecture it opens new machinic possibilities for composition and authorship. When design is considered simply a syntactical process of parts and operations, these can be reconstituted and authored through nature, humans and machines.  

For transformers, an ever growing context-length supports a more nuanced conversation \cite{openai_gpt-4_2023}. Creative processes that improvise, deviate and reformulate established conventions can take advantage of this highly calibrated uncertainty. This is the dialectic between form and possibility or "organised disorder" \cite{eco_open_1989}. Architecture can be ‘read’ in many ways through movement, attention and action that structure aesthetics as language and syntax. Machine learning should not simply seek to transcribe existing universal patterns and probabilities, as in language or images with GPT \cite{openai_gpt-4_2023}, but learn and extend the uncertainties of each circumstance as its own creative act. This is often the case within individual architectural projects where the particulars of a context, materials, community and ideas are overlaid within multimodal assemblages that create a design language unique to that circumstance.   

The Digital Turn promised the rise of “nonstandard seriality” whereby parts no longer need to be identical \cite{carpo_alphabet_2011}, but for architecture unlimited variation is not enough. Endless difference in parts and authorship still needs parsing into contextual architectural outcomes that can be assembled with intelligent probabilities. Machine learning starts to suggest ways to handle and structure this variation, without simply creating variation for its own sake.
  \begin{figure}[h!]
 \centering
  \includegraphics[width=1\textwidth]{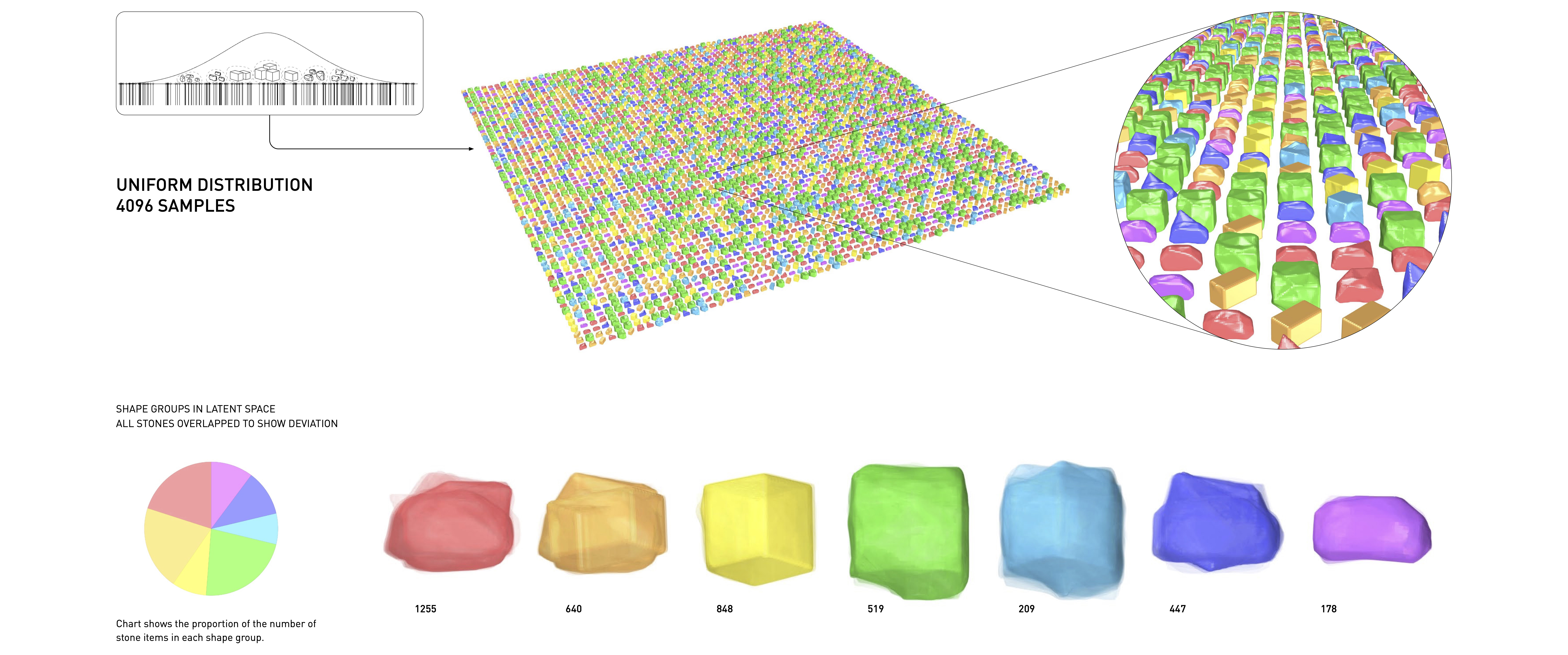}
 \caption{\textit{Probobli Boboli}: a stone training set used to create a model that generates 'natural' forms with precise probabilities.}
 \end{figure}

\textit{\textbf{Probobli Boboli}} \cite{maguire_probobli_2023} reimagines the context of grotto creation in Florence through machine learning. These are spaces where artifice and nature are  blended into a hyper simulated version. A 3D generative model \cite{wu_learning_2017} creates ‘naturalistic’ stone forms indexed to a precise probability distribution of their likely existence (Figure 3).  Parametric scripts select and assemble combinations into alternative complex interlocking structures, each with their own compound probability. This creates an autogenerative work that can generate many alternative design futures, each with a quantified uncertainty from the model (Figure 4).

   \begin{figure}[h!]
 \centering
  \includegraphics[width=1\textwidth]{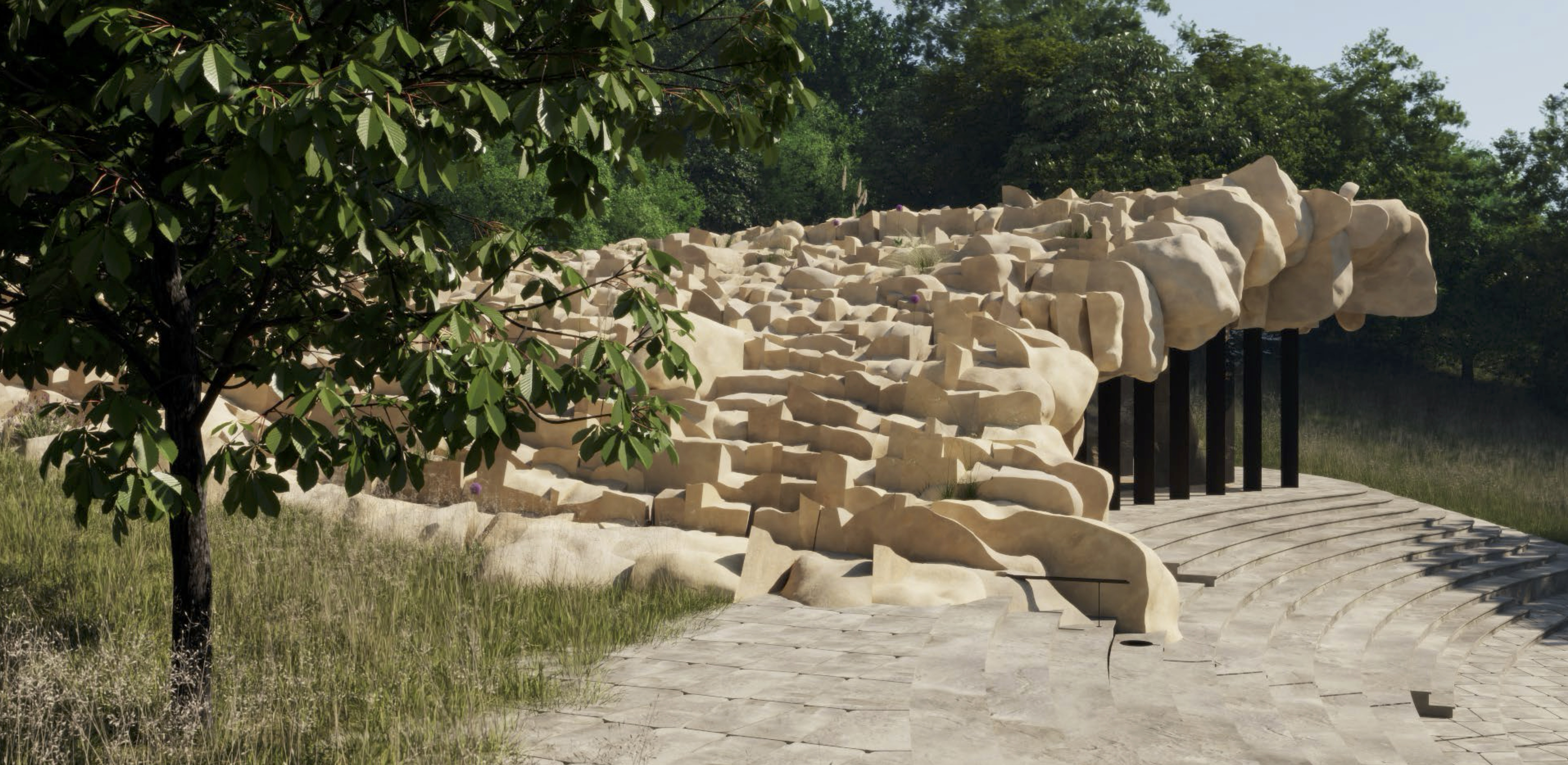}
 \caption{\textit{Probobli Boboli}: Alternative architectural forms are autogenerated according to a recipe of probable stones - analysed, positioned and connected to form surfaces with continuous water drainage paths. See films at \cite{maguire_probobli_2023}.}
 \end{figure}
 
In \textit{\textbf{Towards a Non-Universal Architecture}} \cite{drogeanu_towards_2024} a dataset of cultural gestures in Barcelona is drawn at 1:1 through VR. A 3D generative model \cite{radford_unsupervised_2016} learns the gestural forms indexed to diverse set of participants, and is capable of inferring any specified group. This becomes an allographic instrument that can generate gestures for each hypothetical context (Figure 5). For every site, the ‘community’ of co-authors can be created to a demographic recipe, resulting in an endless auto generative set of architectural designs that are hyper-contextual and co-authored via the machine (Figure 6).
 
     \begin{figure}[h!]
 \centering
  \includegraphics[width=1\textwidth]{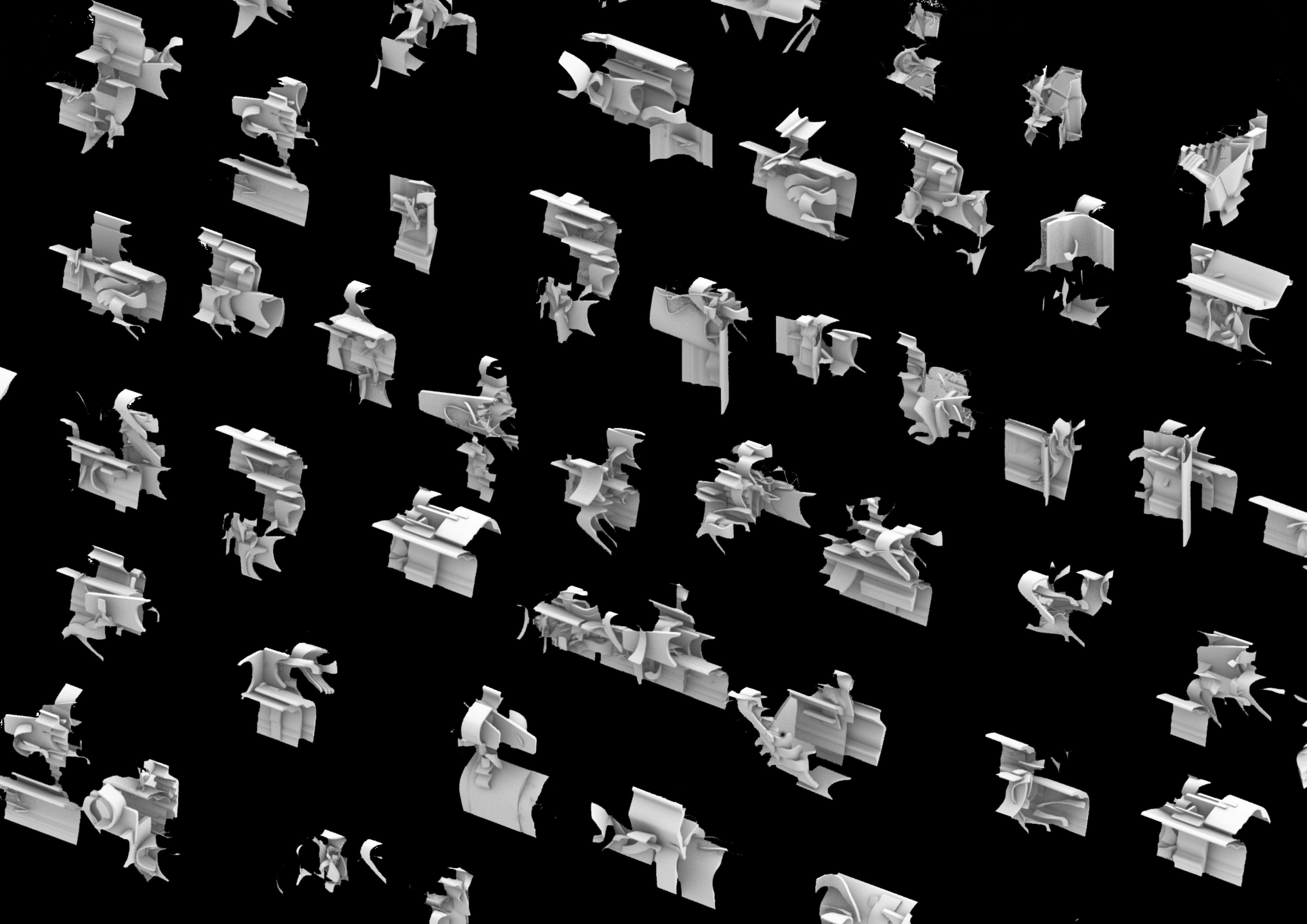}
 \caption{\textit{Towards a Non-Universal Architecture}: autogenerated combinations of gestures that have been generated according to different community recipes, combining VR and GAN interpolation.}
 \end{figure} 
 
     \begin{figure}[h!]
 \centering
  \includegraphics[width=1\textwidth]{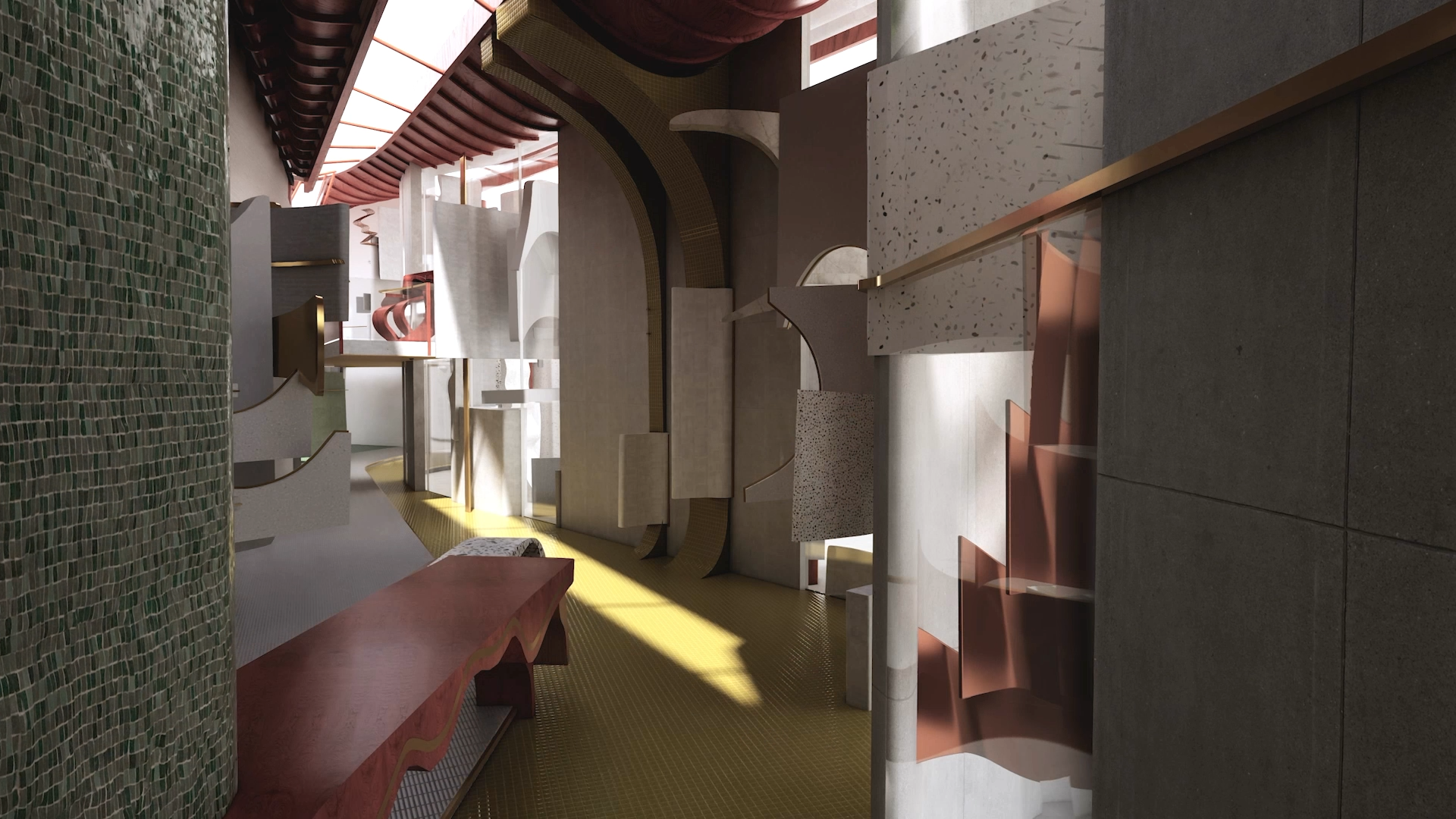}
 \caption{\textit{Towards a Non-Universal Architecture}: This allows an infinite variety of architectural designs to be automatically created that each respond to different locations in Barcelona and different local combinations of community co-authorship. See films at \cite{drogeanu_towards_2024}.}
 \end{figure}  
     \begin{figure}[h!]
 \centering
  \includegraphics[width=1\textwidth]{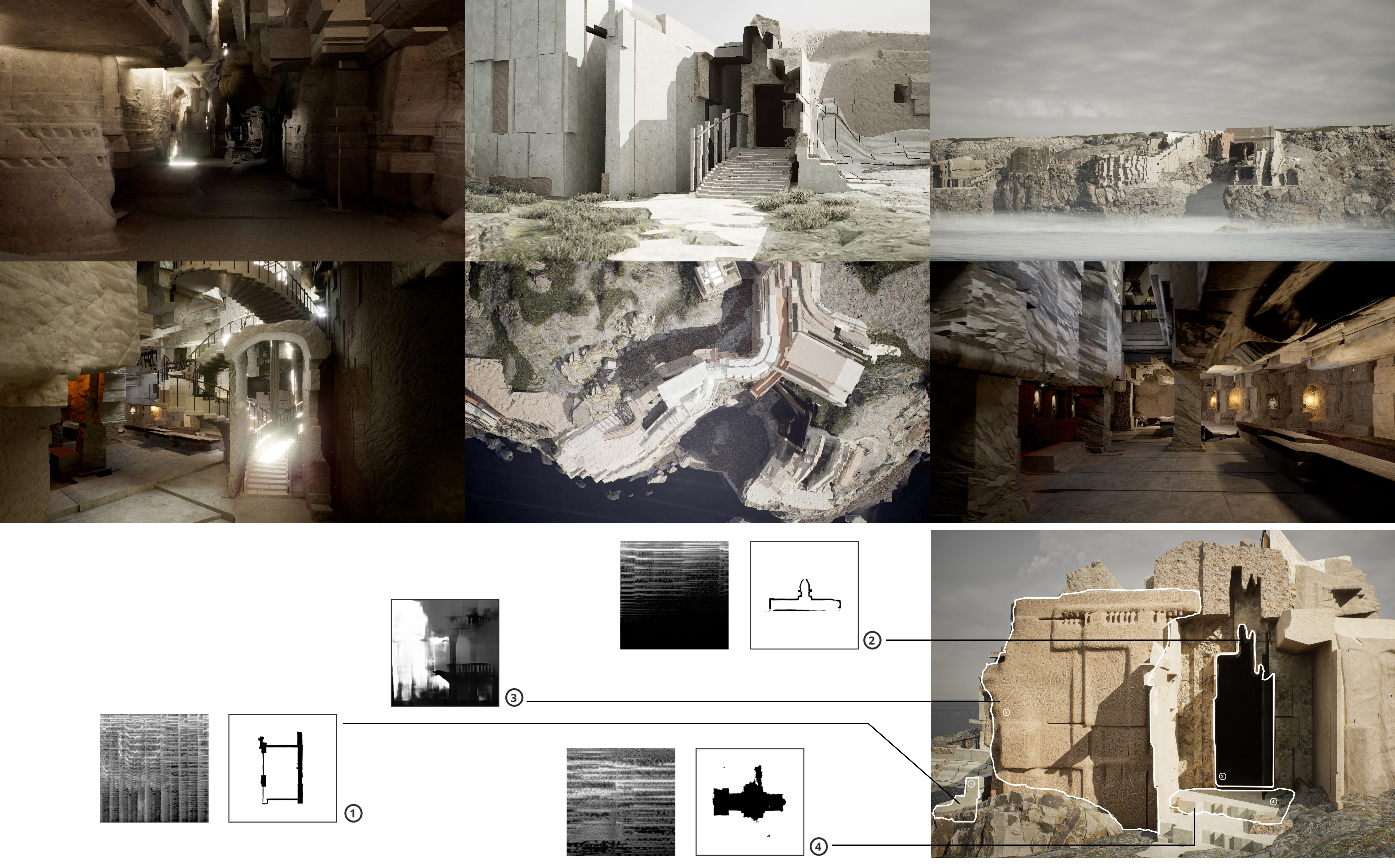}
 \caption{\textit{Crossmodal Compositions}: creating a simultaneous design and musical composition \cite{markevicius_cross-modal_2022}.}
 \end{figure} 
 
\textit{\textbf{Crossmodal Compositions}} \cite{markevicius_cross-modal_2022} (Figure 7) creates a crossmodal interface with conditional GAN that allows the bidirectional translation between drawings and music. This allows a project to create music and architecture in combination. ‘Genres’ of drawing and audio were paired using self-organising maps, describing composition through probability, similar to approaches by Xenakis \cite{xenakis_formalized_1990}.
\newpage
\section{Indeterminacy}

To create with \textit{indeterminacy} is to not just suspend meaning, but test the limits of ontology.   Whilst the word can be regarded as a meta-communicative frame \cite{bateson_steps_1987} that describes how artworks or function are overall kept structurally open (as in \citep{hertzmann_visual_2020}), it is also the sense in which any constituent part has limits and can be \textit{measured}, derived from \textit{determinare}. In complex assemblages such as architecture, individual elements are constantly in negotiation between abstraction and expression. Coded forms of representation and design manipulate the extent of indeterminacy. 

Trained models may appear as a uniformly complete instrument, but in the achievement of a statistically-learnt whole, parts are left incomplete. Machine learning is an incremental process through thousands of training cycles, applying gradual \textit{technogenesis} to allow features to emerge. Just as artworks can have a margin of openness, machines can use “a margin of indeterminacy” that is critical to allow their technical growth \cite{simondon_mode_2017}. In Variational Autoencoders \cite{kingma_auto-encoding_2014} and GAN \cite{goodfellow_generative_2014}, we find input values, and implicit ‘locations’, \textit{probabilised} to match idealised distributions \cite{mackenzie_machine_2017}. Space is not evenly or uniformly trained, but instead achieves its variation and hybridity due to differing degrees of training and determination.

Future creative works should seek to control and calibrate \textit{the amount of} indeterminacy within the latent space through a more direct understanding of its formation in relation to dataset structure. This may be possible with more synthetic data and semi-supervised approaches. Looking beyond an overall Loss calculation, it is in the spaces of the half-trained and half-determinate that we may find the most creative outcomes.

\section{Conclusion}

Margaret Boden suggests human creativity can be defined through three methods: combination, exploration and transformation \cite{boden_creative_2004}. It could be argued \textit{ambiguity}, \textit{uncertainty} and \textit{indeterminacy} respectively align with these approaches. Locating these three terms with more precision might better serve creative uses of AI in future discourse, with meanings that can bridge human ambition and methods of computation.

\textit{Ambiguity} and \textit{Uncertainty} locate the computational upon the world, working through combinatorial, compositional, and statistical patterns that are already established outside the black box and form the template for inference and interpretation. Machine learning as an instrument can hugely amplify and automate but does not necessarily represent a shift from established modes of working. The projects featured show how by taking a multimodal approach we can choose to combine or translate more radically and rapidly. When models are  integrated within a larger creative process it can open up new notions of co-authorship, variation or context - augmenting human creativity rather than emulating it. 

\textit{Indeterminacy} emerges as far more unique quality within machine learning, present within a deeper reading of how latent space forms and achieves a digital ontology from randomness. It suggests there is transductive potential in working between modalities. With further research, this may be of ultimately greater value to the creative process, moving beyond the assumption that training and data must solely emulate established environmental patterns. Datasets might be created, not just curated, in order to exploit the underlying indeterminacy of neural networks and engage with the most 'transformative' creative methods.

\section{Ethical implications}

The example models shown are developed, trained and aligned with very specific design contexts and outputs. They are not considered for a more generalised use with potential unforeseen impacts or biases.  Training data is open source and mostly created through the direct participation of the authors through scanning, modelling, and drawing. This focus on creation through translation does not encourage the mimicry or exploitation of existing creative labour, but instead the augmentation of current creative practice.

\bibliographystyle{plainnat}
\bibliography{testbib3}

\end{document}